\documentclass[10pt]{wsom07}
\usepackage[dvips]{graphicx}
\DeclareGraphicsExtensions{.eps,.ps}
\title{Dynamical Equilibrium, trajectories study in an economical system. The case of the labor market.}
\author{ P. Letr\'emy, M. Cottrell, P. Gaubert, and J. Rynkiewicz\\
 SAMOS-MATISSE-CES, CNRS UMR 8174\\
Paris School of Economy\\
Université Paris 1 Panthéon-Sorbonne\\
90, rue de Tolbiac, 75013 Paris  \\
email: \{patrick.letremy, marie.cottrell, patrice.gaubert, joseph.rynkiewicz\}@univ-paris1.fr\\[1em]
}

\date{}
\begin{document}
\maketitle
%
% Do not include page headings or page numbers
\thispagestyle{empty}
\begin{abstract}
The paper deals with the study of labor market dynamics, and aims
to characterize its equilibriums and
possible trajectories. The theoretical background is the theory of
the segmented labor market. The main idea is that this theory is
well adapted to interpret the observed trajectories, due to the
heterogeneity of the work situations. The Kohonen algorithm is used
to define the segments of the labor market. The trajectories are reconstructed
by means of a non homogeneous Markov model and classified by using a Kohonen algorithm again.

\end{abstract}
\section{Introduction - A classical approach of the labor market}
The analysis proposed here is developed in the framework of the
modern analysis of the labor market. We will first remind briefly
the main features of this approach by comparison with the
classical theory. Our main reference for this presentation is the
Handbook of Labor Economics (\cite{HAND}).

\subsection{Labor market, supply-demand
equilibrium and major determinants.}
 The market is based on a
relation between workers offering their work and firms expressing
a demand, each actor maximizing its function. The equilibrium wage
and the corresponding quantity of labor results from the
comparison between supply and demand. The major characterization
of this solution is productive efficiency. Some strong hypotheses
are behind this result, namely a perfect flexibility of wages and
a global stability of the system, which can be represented by
Walras' "commissaire-priseur". The possibility of persistent
unemployment is explained by a downward rigidity of the wages and
the notion of reservation wage.

\subsection{Global remarks and some critics about the basic
explanation.}
 The representation of the worker's supply behavior,
comparing his reservation wage with the one offered on the market
has from several theoretical points of view, the principal
difficulty being to conciliate a voluntary unemployment with the
supplier rationality. Different approaches in terms of job search
constitute the mainstream in this direction. The initial idea is
to define a rational supplier of labor whose information is
incomplete: during his job search, with diverse costs induced, his
knowledge of the jobs characteristics is imperfect. He has to
choose between, on the one side, the current proposed job with the
actual value of the expected earning, taking account of the
expected stability of this job, and, on the other side, a longer
search for a better job, which means a longer unemployment spell
and higher search costs. This construction makes possible the
introduction of voluntary unemployment and the notion of
reservation wage, its level being defined endogenously in the
model, from the wages distribution and the stability of the
corresponding jobs.

%More recent approaches try to explain more precisely this behavior
%of choice, analyzing more carefully the imperfection of market
%mechanisms, especially asymmetrical information.
%A possible way is
%to introduce uncertainty in the relations existing on the labor
%market and the opposite attitudes of the two categories of agent
%facing risk. This is the analysis developed in the framework of
%implicit contracts theory.

A lot of developments have been made under this approach, leading
to a much more precise representation of behaviors, but also
departing from the initial aim, to explain the mechanism of an
imperfect labor market. As a result, this approach may end, with
the same reasoning, in an equilibrium with unemployment or with
over employment. Among the efforts produced to enhance this
scheme, the idea of incomplete contracts as well as a
macroeconomic approach of indexation may be found.

In the same way, deriving from an individual approach to the
retrieval of a macroeconomic interpretation of disequilibrium,
another main current of thought is that of incentives, with a
special mention to the approach in terms of efficiency wages. An
important result has to be emphasized when compared with the
original project, it is that the theory has moved from an idea of
unemployment as voluntary to that of unemployment as involuntary.
Several ideas can be outlined in this perspective:
\begin{itemize}
\item one of them, very practical, is the approach in terms of
rotation     costs: there is a cost for the firm each time a
worker, once he/she     has received on-job training and acquired
a specific human capital,     moves to another firm in order to
get a better return to his productivity.     To avoid this, the
firm compensates the worker with a wage higher than the market
level, making it more difficult to find a new job better paid.
\item  another one, in a more sociological way, mobilizes the notion
of reciprocal gift;
    it allows for another interpretation of a behavior trying to stabilize
    the work force with the firm: a level of compensation higher than the market
    average is the reciprocate of a level of productivity higher than the
    minimum requirement. Rather than an explanation of the differences
    observed in compensations on a labor market, this approach is
    interesting in the sense that it conceives relationships between
    employers and workers as an adaptive process, resulting from the
    interaction between these agents.
\end{itemize}
This kind of interpretation at the firm level can be developed at
a more general level: in a world of competition, this leads to
some firms paying wages above the market level, inducing in turn
some rigidities in the work force flexibility and a resulting
global unemployment. Numerous empirical studies have shown strong
evidences of this mechanism (see in particular Kruger and Summers,
1988, \cite{KRUG}).

An interesting evolution, in the vein of the present study, is the
one going towards the theory of organizations, that is a different
conception of employer/employee relationship with no reference to
a market and its rules, as it has been presented until now. Again
we find an efficiency wage but it results from a differentiation
process within the firm itself, the firm being viewed as a set of
heterogeneous groups in which the processes of encouragement does
not have to be the same. With this conception, we come closer to
the approach implemented here, with at the beginning the idea that
the labor market is composite, gathering sub-sets using completely
different rules of functioning. These standards models fail to
answer to major questions like the efficiency of policies in that
field. Added to the various weaknesses that have been rapidly
presented above, this leads to look for a different approach of
the labor market as an economic system.

\section{The theory of a segmented labor market: a dynamic approach}
\subsection{The design of the analysis in terms of segmented labor
market} The theory of segmentation, as introduced by Doeringer and
Piore, 1971, \cite{DOER}, and developed later by Piore (see also
Taubman and Wachter, 1986, \cite{TAUB}), defining the principles
of an analysis based on an heterogeneous labor market, is a non
traditional approach.
\subsubsection{\textbf{Heterogeneity of the market}}
- the agents are conditioned by their own history which determines
their choices on the market, currently with the present conditions
existing on this market; ignoring the past of each individual to
keep only the market conditions suppresses an essential component
of the available information (see the US debate on a growing and
persistent labor market instability, in Farber, 1999,
\cite{FARB});

- the agents interact and use the information of past experience
with a progressive learning; put in other words their behavior
results from the past: an adaptive behavior is produced by the
preceding disequilibria; the synthesis realized by Durlauf, 2003,
\cite{DURL} emphasizes the advantages of an approach in terms of
complexity to take into account such behaviors, group effects and
interactions;

- at the same time, this adaptive process concerns the structure
in which the agents interact: this structure is progressively
transformed, and its rules are changing during the whole process.

This means that the traditional approach and its usual tools for
formalization are not adapted to this complexity which constitutes
the foundations of the labor market.
\subsubsection{\textbf{The definition of labor market segments}}
The main distinction is between 1) the primary segment where
institutional rules and organizational practices guaranty both
stability and careers; its main characteristics are rigidities,
the role of social norms, the weight of practices and believes.
This segment cannot be analyzed with a simple comparison between
demand and supply and equilibrium price; 2) the second segment
which is working like a standard market, but which contains a lot
of heterogeneity.
\subsubsection{\textbf{Trajectories}}
This notion is of first importance to understand the system
dynamics. It may represent a set of moves within a given segment
or between two segments (secondary to primary or exit from the
market for an undefined period). Many important questions are
implied by these moves: what is the role of unemployment? Is it a
temporary situation between two segments? Is it better to stay
unemployed than to accept an unskilled job on the secondary
segment, waiting for a good one on the primary? Is gender an
important characteristic in the context? In this perspective, the
trajectories between situations on the labor market cannot be seen
as randomly distributed. A suite of situations have to be
traversed before a given employment may be reached. This order in
the successive situations of employments is called by Piore a
chain of mobility.

This idea of a trajectory or a path followed by workers looking
for a maximization of their objective function is really
attractive, as is the idea of a non homogeneous economy
considering the mechanisms ruling the confrontation between
suppliers of labor and employers. That is the justification for
the identification of trajectories and for the study of the
dynamics of the labor market.
\subsection{Downside of this  analysis.}
The approach in terms of labor market segmentation presents
several important weaknesses. Actually it is a sociological
approach, meaning that the definition of concepts, which would be
necessary to reveal the mechanisms of a segmented economy, are
missing. In place of a statistical measure of the components of
the market, or a set of relations between the analytical
categories determining the dynamics of the system, there is a very
precise description of the types of occupations and skills
constituting the segments. So the theoretical scheme has to be
completed or constructed, using these pertinent observations.
These categories are weakly operative, so it is difficult to
construct an empirical validation of its reality; of course the
construction of models formalizing the implied mechanisms is even
worse. The ambition of the analysis presented after is to help
making a progress in these directions.
\section{The data and the method}
There are few empirical studies in this frame, except to verify
the existence of segments. To tentatively answer the questions
above,  we deal with real data which come from the ``EMPLOI''
(Employment) INSEE (French National Institute for Statistic and
Economy) sample survey. We split our work into several steps.

\begin{itemize}
\item \textbf{Evidence of the reality of the segmentation}\\
The goal is to identify market subsets with very different
characteristics. As we only have the workers data and not the
firms data, we have to identify these subsets by only using the
personal and job characteristics, and not the company policies. We
will consider that the job qualities differentiate the segments,
in terms of career and precariousness (fixed-term or open-ended
contracts). We would try to separate a primary segment (stable and
open-term contracts) from secondary ones (fixed-term
contracts, temporary or seasonal contracts) for example.\\
\item \textbf{Reconstruction of trajectories}\\
We have to observe or to reconstruct the positions occupied over a
long period, to be able to highlight the trajectories and study
the mobility. For example we would like to analyze the role of the
secondary segment with respect to the primary one (is it a simple
transition before first job or before a return to employment after
interruption?). We also want to study the role of the unemployment
situation, with respect to other situations. After an unemployment
period, is it possible to directly return to the primary segment,
or is it usual to go through an alternation of precarious jobs and
unemployment periods?
\end{itemize}
A preliminary study by Gaubert and Cottrell, 1999, \cite{GAUB},
was lead on an American panel over the 1984 to 1992 period,
 for which we had a large number of information about the head
 of households position in the labor market.

The ``EMPLOI'' survey has different characteristics: it is not a
panel since each individual is present only three consecutive
years in the database: this fact has to be overcome as explained
in section 5. This survey is annual and began in 1970. The
questionnaire has changed several times, mainly after each census.
After an important modification in 1990, the questions remained
stable until 2002, and after that, were totally modified and the
survey became quarterly. For that reason, we only consider the
1990-2002 period, that is 13 years.

We only keep the active population, which includes occupied active
peoples, and the unemployed peoples. Due to their specificities,
we remove the farmers. So we can use a more than one million
observations database, represented by several hundreds variables.

There are two kinds of variables: some of them describe the
individual (sex, age, education level, number of children, housing
region, professional activity, and so on); other (more numerous)
describe the individual situation with respect to the labor
market. Most of them are categorical variables (answers to the
inquiry questions), when other ones are numerical (age, salary,
seniority, etc.)

For this part of our work, we used a 3\% sample, randomly drawn in
the total population, that is around 30 000 observations, equally
distributed in the 13 years period. As each individual is observed
three consecutive years on average, we pay attention to keep his
three consecutive observations and remove all those who were
present only once or twice.

The first step is to select the relevant variables to eliminate the
redundancy and to remove those which do not describe the position
of the worker with respect to the labor market. We also reduce the number
of modalities which was excessive in many cases and analyze the non-answers to
see if they correspond to interesting information or to missing data.

The result is that we keep 22 main qualitative variables for a total
number of modalities equal to 99, and on another hand, we consider 28
other variables (qualitative or quantitative) as supplementary variables.
\section{Classification on a Kohonen map, labor market segmentation}
Each observation corresponds to a couple (individual identifier,
year of the observation).

As the main 22 variables are qualitative, we begin by transforming
them into numerical quantitative values without loosing
information. For that we use a Multiple Correspondence Analysis
(MCA) so that each observation is described by its 63 new
coordinates on the factorial axes.
%There is no evidence for
%segmentation from the results of the MCA, so we need to use a
%clustering technique to define the segments.
Due to its visualization and organization properties, we use a
Kohonen map to classify the 30 000 observations (a 8 by 8 grid)
(see \cite{KOHO1}, \cite{COTT}). The number of iterations is 150
000, that is 5 iterations by observation on average, and there is
no pretreatment on the values, since they are already centered and
reduced.

In Fig. 1, we represent the 64 classes and their contents.
The observations (individual, year) are represented
by means of their first 6 factorial components which have the largest variances.
\begin{figure}
\begin{center}
\includegraphics[scale=0.3]{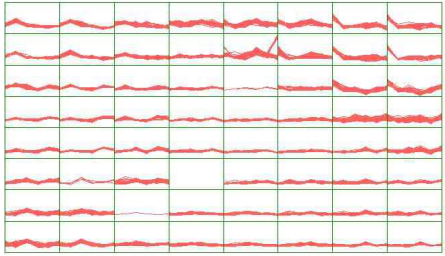}
\caption{The 64 classes and their contents summarized in their 6 first components}
\end{center}
\end{figure}
The observations are regularly distributed over the 64 classes,
only one class is empty, class 30. These 64 classes can be
described by computing elementary statistics of all variables (the
22 main ones, the 28 supplementary ones).

See in Fig. 2, the repartition of the FI variable (FI=1: employed,
FI=2: unemployed), this variable does not
contribute to the classification.
\begin{figure}
\begin{center}
\includegraphics[scale=0.3]{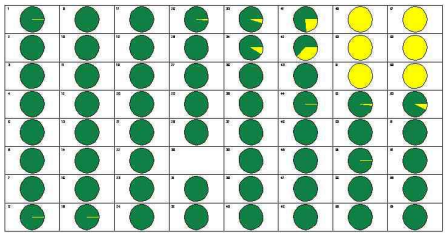}
\caption{The light color indicates the unemployed peoples, they
are all in the upper right corner }
\end{center}
\end{figure}
Fig. 3 shows the repartition of the FIP variable, which indicates
the position of the individual during the previous year. The main
modalities (green-FIP=1 for employed and yellow-FIP=2 for
unemployed) are very stable from one year to the next year, but
unemployed people in classes 45 and 53 were working during the
previous year, the employed people in classes 33 and 34 were
unemployed during the previous year, other modalities (3 to 8)
lead to employment as well as to unemployment.
\begin{figure}
\begin{center}
\includegraphics[scale=0.3]{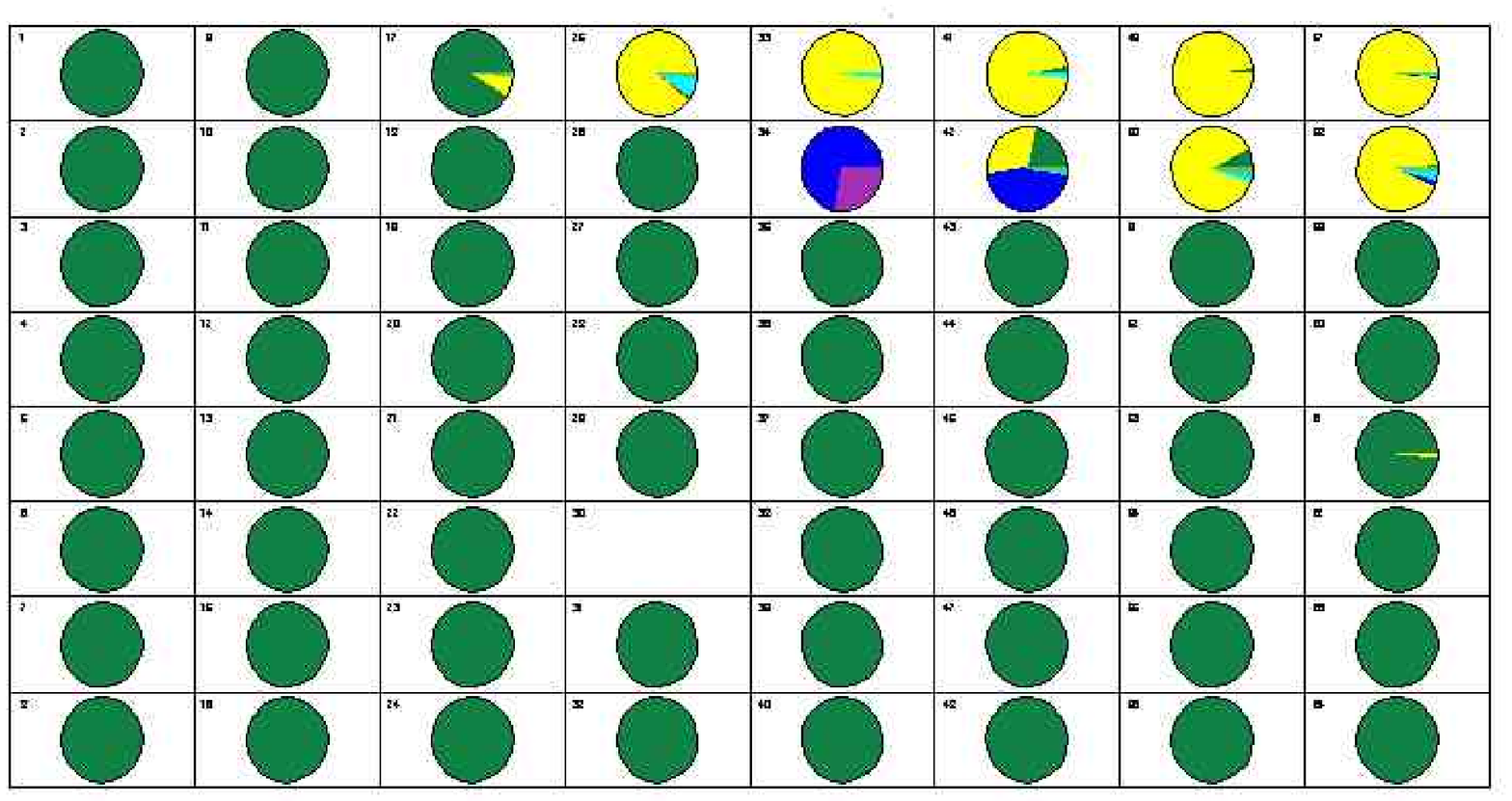}
\caption{Green: employed, yellow: unemployed, blue: student or in training course,
violet: military service, grey and red: retired, removed,
light blue: housewife, light green: other}
\end{center}
\end{figure}
Then, in order to define a small number of segments which will be
easy to identify, we use a Hierarchical Classification ( with Ward
distance, \cite{COTT}, \cite{VESA})to group these 64 classes into
7 super-classes (here called \textbf{segments}), see Fig. 4.
\begin{figure}
\begin{center}
\includegraphics[scale=0.3]{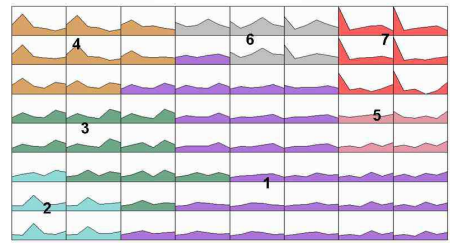}
\caption{The 7 segments}
\end{center}
\end{figure}
Because of the self-organization property of the Kohonen
algorithm, these segments group only neighbor and contiguous
classes. So in this way, we get an accurate and significant
segmentation of the labor market, where we hope to find the
classical segments which were identified by Doeringer and Piore,
1991, \cite{DOER}.

It is easy to identify the 7 segments, from the components of the
code-vectors and also by studying the distribution of the
variables which are not used for the construction of the classes.
Let us give a summary of the description. Segment 1 contains the
workers who have "normal" jobs, clerks, workmen, intermediate
professions (16438 observations). In segment 2 (3029
observations), one finds the open-ended part-time contracts, with
a majority of women and clerks. The managers and intellectuals
form segment 3 (3375 observations). Above it, segment 4 (3006
observations) contains the independent workers on their account
and the craftsmen. Segments 5 and 6 correspond to the situations
close to the unemployment: in segment 5 (1122 observations), we
find the temporary jobs, interim positions, and segment 6 (1309
observations) includes all the peoples who were unemployed during
the previous year. Segment 7 (2660 observations) is exclusively
devoted to unemployed peoples.

After defining the segmentation, it remains to reconstruct the
trajectories among the 7 segments.

\section{The trajectories and their reconstruction}
As each individual is questioned three consecutive years, at least
three states and two transitions are observed for each of them.
The idea is to model these transitions by a Markov chain, in order
to be able to simulate trajectories along the total period and to
compute the limit distribution (given by the probabilities to belong to each segment after a long time).

However, the evolution (see Fig. 5) of the unemployment rate
between 1990 and 2002 shows important variations in the trend and
in the value.
\begin{figure}
\begin{center}
\includegraphics[scale=0.4]{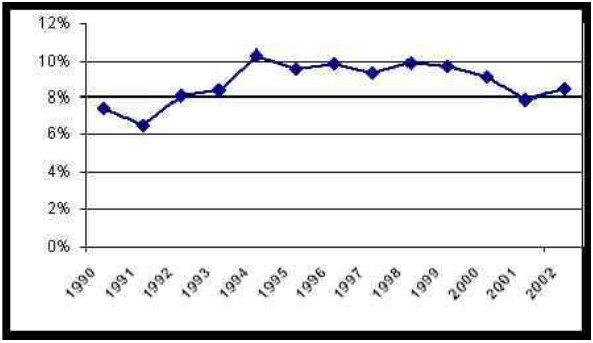}
\caption{The unemployment rate in France along the 1990-2002 period}
\end{center}
\end{figure}
Statistical tests (Maximum Likelihood Tests) prove that it is not
correct to consider an homogeneous Markov chain to model the
changes of segments: there are several significant ruptures over
the period.

To overcome this problem, we define a non homogeneous Markov
chain, where the transition probabilities depend on the current
year. We denote the probability to go in segment $j$ starting from
segment $i$ between year $n$ and year $n+1$ by $p_{ij}^n$. Then
for fixed $i$ and $n$, the vector $( p_{i1}^n, p_{i2}^n, \ldots,
p_{i7}^n)$ defines a probability distribution conditionally to the
fact of belonging to segment $i$ during year $n$.

All parameters $p_{ij}^n$ are estimated from the observed values.

%The precision of these unbiased and consistent estimators would be improved
%if we used the complete database, but it is sufficient for our purpose which
%is to present the methodology.
These estimated conditional distributions can therefore
be used to simulate complete trajectories over the whole period of 13 years.

From the simulated trajectories, we can estimate a "mean" Markov
chain, (see the transition matrix in Table 1) and the "limit"
distribution (Table 2). These estimates are valid as long as the
changes are not too strong, and we can see that the limit values
are very close to the true percentages which are observed in the
database.
\begin{table}
\begin{center}
\begin{tabular}{|c|c|c|c|c|c|c|c|}
\hline
          &  C1   & C2  & C3    &  C4 & C5  &  C6  & C7 \\
\hline
C1   &  \textbf{92.2} &  2.2  &  1.8   &   0.5   &  1.0  &     0.1  &   2.2   \\
\hline
C2  &   \textbf{10.9} &   \textbf{79.4}  &    3.8   &   0.4  &    1.6  &   0.2   &    3.6  \\
\hline
C3  &   9.8     &   3.9  &   \textbf{82.7}  &  1.5   &   0.4 &    0.0   &   1.6          \\
\hline
C4 &    1.9   & 1.0    &   0.4    &  \textbf{95.3}      &    0.2  & 0.0   &    1.2      \\
\hline
C5 &  \textbf{22.0}  &   4.1  &  4.6  &  0.5  &  \textbf{52.7} &   2.4   &   \textbf{13.7} \\
\hline
C6 &    \textbf{28.0} &  \textbf{17.6} &  4.1  &  2.7   &  \textbf{20.4} &  7.6  & \textbf{ 19.7} \\
\hline
C7  &    0.0  &   0.0  &  0.0  &   0.4   & 0.0   &   \textbf{36.8}  &  \textbf{62,8}  \\
\hline
\end{tabular}
\caption{Mean transition matrix, the values are percentages, entry
$(i,j)$ is the probability to go into segment $C_j$ from segment
$C_i$.}
\end{center}
\end{table}
We can see that the exit from segment 7 (unemployed peoples) is
only (with probability 36\%) towards segment 6, what is natural,
since it is the segment of peoples who were without job during the
previous year. But after having spent a year in segment 6, there
are significant probabilities to reach other segments 1, 2, 5.
\begin{table}
\begin{center}
\begin{tabular}{|c|c|c|}
  \hline
  Segment & limit & observed frequency\\
  \hline
Segment 1  &   3.63\%  &3.63\%  \\
Segment 2 &  52.01\%  & 53.13\%  \\
Segment  3  &    9.90\%  &  10.91\% \\
Segment 4  & 11.50\% & 9.79\%   \\
Segment 5  &  12.50\% &  9.72\%  \\
Segment 6   & 3.30\%  & 4.23\%  \\
Segment 7 & 7.80\%  & 8.60\%  \\
\hline
\end{tabular}
\caption{Limit distribution and real values}
\end{center}
\end{table}
We can also draw an individual trajectory, like in Fig. 6.
\begin{figure}
\begin{center}
\includegraphics[scale=0.3]{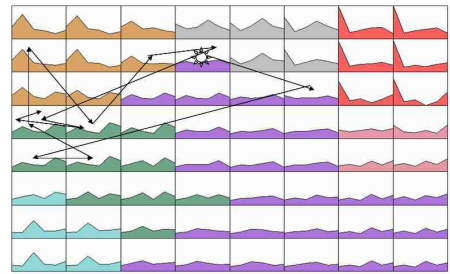}
\caption{The trajectories of one individual who belongs to segment 1
in 1990 and alternates between  to be a manager and an independent worker}
\end{center}
\end{figure}
\section{Classification of trajectories}
Finally, we simulate a large number of trajectories using the
empirical distribution of the individuals in 1990  as initial
state, and the transition probabilities $p_{ij}^n$ to draw the
next state from one year to the next one.

Then it is possible to build a classification of these
trajectories. For this purpose, we use seven one-dimensional
Kohonen maps (strings) with 10 units, to classify all the
trajectories which start from the same initial state. The segment
numbers are reordered to define a kind of scale from the ``best''
situation (segment 1) to the ``worst'' one (segment 7) and we consider
these numbers as real ordered numbers. These classifications
highlight typical behaviors, which remain to be analyzed from an
economical point of view. We use a one-dimensional Kohonen to get an "ordered" sequence of possible trajectories.

In Figure 7, the code vectors of the 7 Kohonen maps are displayed.
The left column indicates the percentages of the individuals
according to their situation in 1990. In lower part of each
10-units map, the frequencies of each class are written.
\begin{figure}
\begin{center}
\includegraphics[scale=0.4]{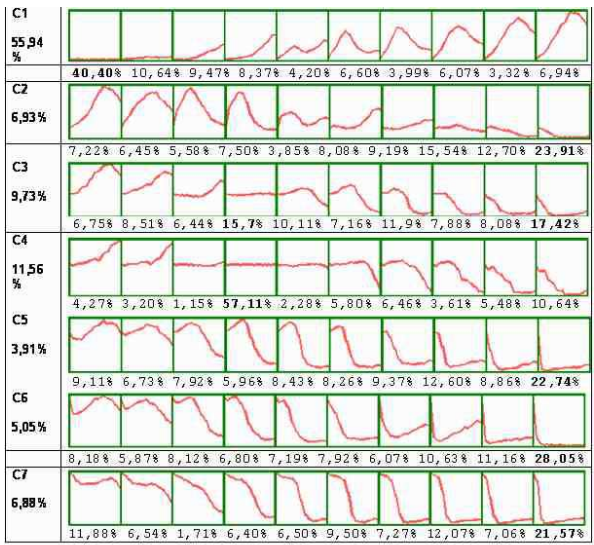}
\caption{The classes of the trajectories according to the initial
State of the worker}
\end{center}
\end{figure}
We observe that most of workers in ``normal situation'' in 1990
remain in this ``normal'' situation, that segments 3 and 4 are
stable, that even if most of unemployed peoples remain close to
segment 7 or 6, a not negligible proportion goes to segment 1 (finds a job), and so on.

The study of these trajectories allows us to analyze in a very
detailed way typical behaviors which were not clear by considering
only the segmentation.
\section{Conclusion and perspectives}
From the SOM point of view, the more interesting result is that we
extend its traditional domain of application to economical theory
and to dynamical behavior models. Future works include a more
precise analysis of the meaning of each typical trajectory. This
analysis has to be done in relation with the personal profiles of
the individuals. Up to now, we have not used all the variables
which are present in the survey. In particular the variables which
do not directly describe the situation with respect to the labor
market were not considered for the segmentation, but could explain
the heterogeneity of the trajectories from a worker to another
one.

\end{document}